\documentclass[
 reprint,
superscriptaddress,
 amsmath,amssymb,
 aps, nofootinbib, preprintnumbers
]{revtex4-2}

\usepackage{graphicx}
\usepackage{dcolumn}
\usepackage{bm}
\usepackage{float}
\usepackage{epsfig}
\usepackage{amsfonts}
\usepackage{amssymb}
\usepackage{amsmath}
\usepackage{mathtools}
\usepackage{multirow}
\usepackage{lipsum}
\usepackage{physics}

\usepackage{rsfso}
\usepackage{mathrsfs}
\usepackage{booktabs}
\usepackage[normalem]{ulem}
\usepackage{placeins}
\usepackage{xcolor}
\usepackage{MnSymbol,wasysym}
\DeclareMathAlphabet{\mathcal}{OMS}{cmsy}{m}{n}
\usepackage[hidelinks]{hyperref}
\hypersetup{colorlinks=false,breaklinks=true}

\def\D{\mathcal{D}}
\usepackage{cancel}

\newcommand{\bk}{\boldsymbol{k}}

\newcommand{\bq}{\boldsymbol{q}}
\newcommand{\bx}{\boldsymbol{x}}

\newcommand{\de}{{\rm d}}
\def\bea{\begin{eqnarray}}
\def\eea{\end{eqnarray}}
\def\be{\begin{equation}}
\def\ee{\end{equation}}
\def\ba{\begin{array}}
\def\ea{\end{array}}
\def\nn{\nonumber}

\usepackage{tikz}
\usetikzlibrary{calc}
\usetikzlibrary{arrows,patterns}

\begin{document}

\title{A note on loop resummation in de Sitter spacetime with the wavefunction of the universe approach}

\author{Javier Huenupi}
 \email{javier.huenupi@ug.uchile.cl}
  \affiliation{Departamento de F\'isica, Facultad de Ciencias F\'isicas y Matem\'aticas, Universidad de Chile, Santiago, Chile}
\author{Ellie Hughes}
 \email{ellieh@mit.edu}
  \affiliation{Center for Theoretical Physics, Massachusetts Institute of Technology, Cambridge, Massachusetts 02139, USA}
\author{Gonzalo A. Palma}
 \email{gpalmaquilod@ing.uchile.cl}
  \affiliation{Departamento de F\'isica, Facultad de Ciencias F\'isicas y Matem\'aticas, Universidad de Chile, Santiago, Chile}
\author{Spyros Sypsas}
 \email{s.sypsas@gmail.com}
  \affiliation{High Energy Physics Research Unit, Faculty of Science, Chulalongkorn University, Bangkok 10330, Thailand}
  \affiliation{National Astronomical Research Institute of Thailand, Don Kaeo, Mae Rim, Chiang Mai 50180, Thailand}

\preprint{MIT-CTP/5809}

\date{March 24, 2025} 

\begin{abstract}

We analyze the computation of $n$-point correlation functions in de Sitter spacetime, including loop corrections, using the wavefunction of the universe approach. This method consists of two stages employing distinct Feynman rules. First, one must compute the wavefunction coefficients using interactions as vertices. Then, in the second stage, one computes correlation functions using wavefunction coefficients as vertices. For massless fields, loop corrections in the first stage are free of infrared (IR) divergences, which leads to the question of how this matches the well-known IR behavior of correlators obtained via other methods. By considering a scalar field with an arbitrary potential, we compute $n$-point correlation functions to first order in the potential but to all orders in loops. We find that, although loop integrals in the first stage are indeed IR convergent, the second procedure reintroduces the IR divergence. We discuss how this induces renormalization of the interaction potential such that the final result combining both steps exactly matches the form of $n$-point functions previously calculated with other methods.

\end{abstract}

\maketitle

\usetikzlibrary {patterns.meta}
\tikzdeclarepattern{
  name=mylines,
  parameters={
      \pgfkeysvalueof{/pgf/pattern keys/size},
      \pgfkeysvalueof{/pgf/pattern keys/angle},
      \pgfkeysvalueof{/pgf/pattern keys/line width},
  },
  bounding box={
    (0,-0.5*\pgfkeysvalueof{/pgf/pattern keys/line width}) and
    (\pgfkeysvalueof{/pgf/pattern keys/size},
0.5*\pgfkeysvalueof{/pgf/pattern keys/line width})},
  tile size={(\pgfkeysvalueof{/pgf/pattern keys/size},
\pgfkeysvalueof{/pgf/pattern keys/size})},
  tile transformation={rotate=\pgfkeysvalueof{/pgf/pattern keys/angle}},
  defaults={
    size/.initial=5pt,
    angle/.initial=45,
    line width/.initial=.4pt,
  },
  code={
      \draw [line width=\pgfkeysvalueof{/pgf/pattern keys/line width}]
        (0,0) -- (\pgfkeysvalueof{/pgf/pattern keys/size},0);
  },
}

\section{Introduction} \label{sec:introduction}

Understanding the behavior of correlation functions of light fields in de Sitter spacetime is crucial for exploring quantum field theory in cosmological contexts~\cite{Vilenkin:1982wt,Linde:1982uu,Starobinsky:1982ee,Tsamis:1994ca,Tsamis:2005hd,Seery:2010kh,Burgess:2010dd,Arai:2011dd,Tanaka:2013caa,Baumgart:2019clc,Mirbabayi:2019qtx,Senatore:2009cf,Gorbenko:2019rza,Palma:2023idj,Palma:2023uwo,Cespedes:2023aal,Negro:2024bbf,Kamenshchik:2024ybm,Chen:2023bji}. Its far-reaching implications not only affect the status of de Sitter space itself as a stable gravitational background~\cite{Ford:1984hs,Antoniadis:1985pj,Polyakov:2012uc} but also cosmological phenomenology~\cite{Bartolo:2007ti,Pattison:2017mbe,Chen:2018brw,Chen:2018uul,Panagopoulos:2019ail}. Indeed, the observed cosmic web reflects the initial conditions that led to its formation, which are generated dynamically during an early de Sitter evolutionary stage: cosmic inflation. 

There exist several methods to compute such correlation functions. In flat-space QFT, one defines asymptotically free states which allow for a perturbative computation of in-out amplitudes encoding the probability of an initial condition to evolve into a final state. In curved spacetime, due to spontaneous particle production induced by curvature~\cite{Parker:1968mv}, defining asymptotic states is subtle. What one can safely do
is to define quantum states momentarily. To adapt to this feature, the flat-space operator formalism generalizes to the in-in formalism~\cite{Maldacena:2002vr,Weinberg:2005vy,Adshead:2009cb}, where the main object of interest becomes the equal-time $n$-point function, which correlates fixed initial conditions (``in'' states) as they evolve towards the future. 

In path integral language, switching to in-in correlators schematically corresponds to a folding of the temporal contour with respect to the flat-space path integral. The result is a path integration along the Schwinger-Keldysh (SK) contour~\cite{Calzetta:1986ey,Chen:2017ryl}, which essentially captures the correlation of states defined in the asymptotic past, at a specific moment in their evolution corresponding to the folding point of the contour. The partition function then serves as a generating function of the aforementioned equal-time correlators. 

The third way of computing correlators, and the method employed herein, is the wavefunction of the universe (WFU) formalism~\cite{Hartle:1983ai,Maldacena:2002vr,Anninos:2014lwa,Pajer:2020wxk,Arkani-Hamed:2018kmz,Goodhew:2024eup}. This approach is quite intuitive in time-dependent spacetimes, such as cosmological de Sitter, which provides a fairly accurate description of the inflationary background. Having a 3+1 dimensional spacetime foliation in mind and given that we are interested in correlators at a fixed time-slice, as per common practice, we distinguish between \emph{boundary} objects depending only on the time of interest, and \emph{bulk} objects that depend on the whole history.

In a nutshell, the wavefunction of the universe dictates the projection of a boundary quantum state onto the basis of bulk-field eigenstates, which, in turn, encodes the probability that a given bulk configuration ends up at a specific boundary state. 
Due to this bulk-boundary split, it is no accident that this formalism has been instrumental in the development of the cosmological bootstrap program~\cite{Arkani-Hamed:2018kmz,Pajer:2020wxk}, which seeks to derive properties of boundary observables from fundamental principles like locality, unitarity, and causality, without explicit reference to their bulk dynamics.

In this work, we will limit ourselves to delineating the exact correspondence between the WFU and the in-in or SK formalisms for a scalar field in de Sitter space with arbitrary self-interactions. Our calculations are exact to linear order in the potential. The nontrivial character arises due to the generality of the interaction, which, in perturbation theory, translates to an infinite number of vertices. Consequently, each $n$-point correlator receives contributions from all the vertices of order $m\geq n$, which are organized in an infinite series of corrections to the bare couplings. Diagrammatically, these corrections can be represented as \emph{daisy} loops (carrying no external momentum) dressing the interaction vertices~\cite{Chen:2018brw}. 
The resulting Witten diagrams are the simplest nontrivial diagrams capturing most of the subtleties arising in de Sitter dynamics that can be analytically computed; for example, just like higher order loops, daisy loops lead to divergences which have to be dealt with.  
The goal of this note is to show exactly how the renormalization of these diagrams operates in the WFU approach. 
 
\section{Renormalization of Schwinger-Keldysh correlators}
\label{sec:SK}

To set the stage, let us first review the renormalization of $n$-point SK correlation functions~\cite{Chen:2018brw,Palma:2023idj,Huenupi:2024ksc}. (The in-in computation is equivalent.) We parametrize the Poincar\'e patch of de Sitter space (dS) via the Friedmann-Lemaître-Robertson-Walker (FLRW) line element: $\dd s^2 = a(\tau)^2\left(-\dd \tau^2 + \dd x^2 \right)$, with $\tau\in(-\infty,0)$ the conformal time. The scale factor is given by $a(\tau)=1/|H\tau|$, with the Hubble rate, $H$, setting the de Sitter radius. Our scalar field theory is then governed by the following action: 
\be
S[\phi] = \int \! \dd[3] x   \dd\tau\, a^4 \bigg[ \frac{\dot \phi^2}{2a^2} - \frac{( \nabla \phi )^2}{2 a^2}  - {\mathcal V}_{\rm b}(\phi) \bigg] , \label{intro:action}
\ee
where an overdot denotes a derivative with respect to conformal time. 

Our only assumptions are that 1) the arbitrary bare potential ${\mathcal V}_{\rm b}(\phi)$ is analytic, that is, it equals its Taylor series at any point, and 2) it is subleading compared to the Hubble rate, such that the probe (or spectator) approximation holds.\footnote{Upon coupling such spectators to the inflaton in some model-dependent way, their statistics can be directly transferred to the observable curvature-perturbation distribution, rendering their study also phenomenologically relevant.} A prime example of this sort of dynamics is the case of axions during inflation, with a potential given by ${\cal V}(\phi) \propto 1-\cos(\phi/f)$. Note that such potentials are bounded and can thus be accurately treated within the perturbative scheme discussed here.

Given the action in Eq.~\eqref{intro:action}, we may first quantize the free theory in the usual manner. For simplicity, we choose to perturb around the massless theory with Bunch-Davies initial conditions, utilizing the positive energy dS mode function:
\begin{equation} \label{ds-mode+}
f^+_k(\tau) = \frac{H}{\sqrt{2k^3}}(1 + ik\tau)e^{-ik\tau}.
\end{equation}
Given the linear solution, we can compute all four SK propagators, which, due to the structure of the diagrammatics, can be written in terms of the following combination:
\be
g(k, \tau_1,\tau_2) = f_k^+(\tau_1) f_k^{-}(\tau_2) , \label{G-uu}
\ee  
with $f_k^-$ the complex conjugate of Eq.~\eqref{ds-mode+}.

An important quantity permeating perturbation theory is the second cumulant, defined as
\be \label{variance-0}
\sigma_{\rm tot}^2 \equiv \int_{\bk} \, g (k, \tau,\tau), 
\ee
where $\int_{\bk}\equiv\int\frac{\dd[3]k}{(2\pi)^3}$.
To deal with $\sigma_{\rm tot}^2$, it is convenient to perform a change of integration variables from comoving momentum $k$ to physical momentum $p \equiv k / a(\tau) = - H k \tau$. After this step, $\sigma_{\rm tot}^2$ becomes manifestly time independent:
\be
\sigma_{\rm tot}^2 = \frac{H^2}{4 \pi^2} \int_{0}^{\infty} \frac{\de p}{p} \left( 1 + \frac{p^2}{H^2} \right) . \label{variance-0-1}
\ee
This variance is constant and divergent in both the ultraviolet (UV) and the infrared (IR) limits.\footnote{The IR divergence is absent in the case of a free massive scalar since the mass provides a physical IR cutoff~\cite{Arai:2011dd}.}

To proceed with the interactions in a perturbative manner, we expand the bare potential as
\be \label{V-bare-T}
\mathcal{V}_{\rm b}(\phi) = \sum_{n=2}^\infty \frac{\lambda_n^{\rm b}}{n!} \phi^n.
\ee
With this expansion, the SK formalism tells us that each $n$-point function is sourced by an infinite series of $(n+2L)$-legged vertices proportional to $\lambda_{n+2L}$, in such a way that $2L$ of the available legs end up forming closed loops. The final result is a tree-level $n$-point correlation function proportional to an effective coupling $\lambda^{\rm obs}_n$ given by
$
\big \langle \phi_{ \bk_1 \cdots \bk_n} \big \rangle_c  = (2 \pi)^3 \delta^{(3)} \left(\sum_i \bk_i\right) \big \langle \phi_{ \bk_1 \cdots \bk_n} \big \rangle ',
$
with 
\be \label{npt'}  
\!\!\!\! \big \langle \phi_{ \bk_1 \cdots \bk_n} \big \rangle ' \!\! =  \!  \frac{\lambda_n^{\rm obs}}{H^4} 2\,{\rm Im} \Bigg\{ \! \int^{\tau}_{- \infty} \frac{\de \bar\tau}{\bar\tau^4} g (k_1,\tau, \bar\tau)  \cdots g (k_n,\tau, \bar\tau) \Bigg\} .
\ee
In the previous expression, the effective coefficients $\lambda^{\rm obs}_n$ are given by~\cite{Chen:2018brw,Palma:2023idj,Huenupi:2024ksc}
\be
 \lambda^{\rm obs}_n =  \sum_{L=0}^{\infty}   \frac{ \lambda^{\rm b}_{n+2L} }{L!} \left[ \frac{\sigma_{\rm tot}^2}{2} \right]^L  . \label{lambda-ren}
\ee
The label $L$ counts the number of loops participating in a diagram with $n$ external legs sourced by the coefficient $\lambda_{n +2 L}$.

Because of Eq.~\eqref{variance-0-1}, the effective coupling $\lambda_{n}^{\rm obs}$ in Eq.~\eqref{lambda-ren} is formally divergent. However, as shown in Ref.~\cite{Huenupi:2024ksc}, to linear order in the interaction potential (i.e., for single-vertex diagrams), the theory in Eq.~\eqref{intro:action} can be renormalized in the sense that each bare coupling constant $\lambda_n^{\rm b}$ induces a sequence of counterterms which removes the divergent induced by $\sigma_{\rm tot}^2$, order by order in terms of loops,
leading to a finite \emph{observable} coupling. More to the point, one can write $\lambda_{n}^{\rm b} = \lambda_n + \sum_{L=1}^{\infty} \lambda_n^{(L)}$ where $\lambda_n$ is the renormalized coupling and the coefficients $\lambda_n^{(L)}$ are counterterms needed to eliminate divergences order by order. One can then write $\sigma_{\rm tot}^2 = \sigma^2 + \sigma_{\infty}^2$, and tune the counterterms to get rid of the divergent part $\sigma_{\infty}^2$. In doing so, one finally obtains
\be
\lambda^{\rm obs}_n =  \sum_{L=0}^{\infty}   \frac{ \lambda_{n+2L} }{L!} \left[ \frac{\sigma^2}{2} \right]^L  , \label{lambda-ren-0}
\ee
which is explicitly finite. Of course, the splitting $\sigma_{\rm tot}^2 = \sigma^2 + \sigma_{\infty}^2$ is arbitrary, but the definition of $\lambda^{\rm obs}_n$ is not. This implies that the renormalized coupling $\lambda_{n}$ must run with $\sigma^2$ in such a way that $\lambda^{\rm obs}_n$ remains independent of the splitting, inducing a renormalization group equation for each coupling.

Interestingly, the observable couplings $\lambda_{n}^{\rm obs}$ defined in Eq.~\eqref{lambda-ren} [or, alternatively, in Eq.~\eqref{lambda-ren-0}] can be identified as the Taylor coefficients of an observable potential $\mathcal V_{\rm obs} (\phi)$: 
\be \label{Vobs-def}
\mathcal V_{\rm obs} (\phi) \equiv \sum_{n} \frac{\lambda_n^{\rm obs}}{n!} \phi^n .
\ee
Thanks to Eq.~\eqref{lambda-ren}, one can verify that $\mathcal V_{\rm obs} (\phi) $ and the bare potential $\mathcal V_{\rm b} (\phi) $ are related through a Weierstrass transform, taking the form~\cite{Chen:2018brw,Palma:2023idj,Huenupi:2024ksc}
\be \label{Vobs-W}
\mathcal V_{\rm obs} (\phi) = e^{\frac{1}{2} \sigma_{\rm tot}^2 \partial^2_\phi}  \mathcal V_{\rm b} (\phi) .
\ee
Alternatively, one may also define a renormalized potential $\mathcal V (\phi) = \sum_{n} \frac{\lambda_n}{n!} \phi^n $ such that the coefficients $\lambda_n$ are those appearing in Eq.~\eqref{lambda-ren-0}. Then, again, one can derive the following identities involving Wierstrass transforms: $\mathcal V_{\rm obs} (\phi) \equiv e^{\frac{1}{2} \sigma^2 \partial^2_\phi}  \mathcal V (\phi) $ and $\mathcal V (\phi) = e^{\frac{1}{2} \sigma_{\rm \infty}^2 \partial^2_\phi}  \mathcal V_b (\phi)$.

In what follows, we first extend previous results~\cite{Lee:2023jby,Creminelli2024} to the case of infinite loops dressing $n$-point correlators: We demonstrate that the WFU and SK formalisms are equivalent for the aforementioned particular set of diagrams which add up to the correlator in Eq.~\eqref{npt'}. In addition, we show how the observable couplings of Eq.~\eqref{lambda-ren} appear in the WFU approach. Contrary to the SK computation, it is known that, here, the IR cutoff has an obscure role~\cite{Anninos:2014lwa,Creminelli2024}. 
Indeed, when one computes the wavefunction coefficients in perturbation theory one encounters no need for an infrared regulator since the resummation of bulk loops involves an IR-finite variance. However, in the second step where one computes correlation functions given these wavefunction coefficients, the IR divergence reappears through boundary loops. In particular, we demonstrate that, as observed in Refs.~\cite{Lee:2023jby,Creminelli2024} for a single loop, bulk and boundary loops combine in tandem to yield Eq.~\eqref{lambda-ren}, which resums an infinite number of loops.

\section{Wavefunction of the universe}

We refer the reader to Appendix A of Ref.~\cite{Goodhew:2020hob} for a concise pedagogical introduction to the formalism.
Since the bulk-boundary language will now be important, we need to differentiate the notation between the bulk field and the boundary configuration. 
We will denote by $\tau$ the location of the boundary. 

We wish to compute the correlators in Eq.~\eqref{npt'}. Since these are boundary objects, to do so, the very first thing we need is an orthonormal basis of boundary field-eigenstates. 
To that end, we define the bulk-field operator $\Phi(\bar\tau,\bx)$, whose spectrum spans all the bulk field-configurations ending on the same boundary profile
$\phi(\bx) \equiv \phi(\tau,\bx)$:
\begin{equation}
\Phi(\bar\tau,\bx)\ket{\phi(\bx)} = \phi(\bx)\ket{\phi(\bx)}.
\end{equation}
A boundary state $\ket{\Psi}$ can then be constructed as a linear combination of the field-eigestates $\ket{\phi}$: 
\be 
\ket{\Psi} = \int\limits_{\substack{\Phi(\tau) = \phi \\ \Phi(-\infty) = 0}}\!\! \mathcal{D}\Phi\, \Psi[\Phi] \ket{\phi}.
\ee   
In the field representation, the wavefunction then reads
\begin{equation}\label{eq:wfu-def}
\Psi[\phi] \equiv \bra{\phi(\bx)}\ket{\Psi} .
\end{equation}

The correlators in Eq.~\eqref{npt'} are computed in the Bunch-Davies state, which provides a vacuum state at past infinity. We thus need to choose $\ket{\Psi}$ in Eq.~\eqref{eq:wfu-def} to be the vacuum state at $\tau=-\infty$ and then use 
the evolution operator to evolve it towards the boundary. 
After some algebra~\cite{Goodhew:2020hob}, the result is 
\be \label{Psi-PI}
\Psi[\phi] \propto \int\limits_{\substack{\Phi(\tau) = \phi \\ \Phi(-\infty) = 0}}\mathcal{D}\Phi \, e^{iS[\Phi]}.
\ee

Via a saddle-point approximation, this object can be evaluated in a nonperturbative manner~\cite{Celoria:2021vjw}. However, here we will be interested in a perturbative expansion around the free theory. To implement such a scheme, it is convenient to expand the wavefunction in a series as
\begin{equation}\label{eq:series-expansion-WFU}
\Psi[\phi] \! =  \!\mathcal{N}\!\exp\!\Bigg\{\! \sum_{m=2}^\infty\frac{1}{m!}\int_{\bk_1,\dotsc,\bk_m}\!\!\!\!\!\psi_m(\bk_1,\dotsc,\bk_m)\phi_{\bk_1}\dotsb\; \phi_{\bk_m}\!\Bigg\},
\end{equation}
where $\cal N$ is a normalization factor and
\begin{equation}
\!\!\!\psi_m(\bk_1,\dotsc,\bk_m) = (2\pi)^3 \delta^{(3)}\left(\sum_i \bk_i\right) \psi_m'(\bk_1,\dotsc,\bk_m),
\end{equation}
are the Fourier-space WFU coefficients.
These can then be computed 
by taking functional derivatives of the wavefunction:
\begin{flalign}
\psi_m(\bk_1,\dotsc,\bk_m) &= \frac{1}{\Psi[0]}\frac{\delta \Psi[\phi]}{\delta\phi_{\bk_1}\dotsb\, \delta\phi_{\bk_m}} \notag \\
&= \frac{i^m}{\Psi[0]}\int\mathcal{D}\Phi \, \Pi_{-\bk_1}\dotsb\; \Pi_{-\bk_m}\exp(iS[\Phi]),
\label{eq:psi_n-formula}
\end{flalign}
where $\Pi_{\bk} \equiv a^2\partial_{\bar{\tau}}\Phi_{\bk}(\bar{\tau})\Big|_{\bar\tau = \tau}$ is the conjugate momentum evaluated at the boundary.

The computation proceeds by expanding the exponential to the desired order and contracting bulk fields among themselves and with boundary fields, via the bulk-to-bulk and bulk-to-boundary propagators $G(k_1;\tau_1,\tau_2)$ and $K(k,\tau_1)$, respectively:
\begin{equation}
\expval{\Phi_{\bk_1}(\tau_1)\Phi_{\bk_2}(\tau_2)} = G(k_1;\tau_1,\tau_2)(2\pi)^3\delta^{(3)}(\bk_1 + \bk_2),
\end{equation}
and
\begin{equation}
\expval{\Pi_{\bk}(\tau)\Phi_{\bk_1}(\tau_1)} = -iK(k,\tau_1)(2\pi)^3\delta^{(3)}(\bk + \bk_1).
\end{equation}
Explicitly, these are given by
\begin{flalign}
G(k;\tau_1,\tau_2) &= f^+_k(\tau_1)f^-_k(\tau_2)\theta(\tau_1 - \tau_2) \notag \\
& + f_k^+(\tau_2)f^-_k(\tau_1)\theta(\tau_2 - \tau_1) \notag \\
& - \frac{f_k^+(\tau)}{f^-_k(\tau)}f^-_k(\tau_1)f^-_k(\tau_2),
\label{eq:G-gral-exp}
\end{flalign}
and
\begin{equation}\label{eq:K-gral-exp}
K(k,\tau_1) = \frac{f^-_k(\tau_1)}{f^-_k(\tau)},
\end{equation}
with the latter following from the definition of the conjugate momentum, which implies that $a^2\partial_{\bar{\tau}} G(k;\bar{\tau},\tau_1)\Big|_{\bar\tau = \tau} = -iK(k,\tau_1)$.

Note that, in contrast with Section~\ref{sec:SK}, here, we need to choose the negative-frequency mode function $f^-_k$ in order to satisfy the initial condition
$
\Phi(\bx,-\infty) = 0
$ appearing in the path integral in Eq.~\eqref{Psi-PI}.
For a massless field, this is just the complex conjugate of Eq.~\eqref{ds-mode+}.

\section{Bulk-loop-resumed wavefunction coefficients}\label{subsec:WFU-1-order}

We will compute the wavefunction coefficients $\psi_m$ to first order in the interaction potential, using standard diagrammatical rules. An $n$-legged vertex, characterizing a monomial of order $n$ in the Taylor expansion of the interaction potential, $\mathcal{V}(\Phi) \supset \frac{\lambda_n}{n!}\Phi^n$, corresponds to an integral of the following form:
\def\nvertexb{\tikz[baseline=-0.6ex,scale=1.8, every node/.style={scale=1.4}]{
\coordinate (v1) at (0ex,0ex);
\coordinate (phi1) at (-3ex,2ex);
\coordinate (phi2) at (-4ex,0ex);
\coordinate (phi3) at (-3ex,-2ex);
\coordinate (phi4) at (3ex,2ex);
\coordinate (phi5) at (4ex,0ex);
\coordinate (phi6) at (3ex,-2ex);
\draw[thick] (v1) -- (phi1);
\draw[thick] (v1) -- (phi2);
\draw[thick] (v1) -- (phi3);
\draw[thick] (v1) -- (phi4);
\draw[thick] (v1) -- (phi5);
\draw[thick] (v1) -- (phi6);
\filldraw[color=black, fill=black, thick] (v1) circle (0.1ex);
\node[anchor=south] at ($(v1)+(0,-2.5ex)$) {\scriptsize{$\bar\tau$}};
\node[anchor=south] at ($(v1)+(0,+1.0ex)$) {\scriptsize{$\cdots$}};
\node[anchor=south] at ($(v1)+(0,+2.0ex)$) {\scriptsize{$n\;\rm{legs}$}};
}
}
\def\internalpropagatorWFU{\tikz[baseline=-0.6ex,scale=1.8, every node/.style={scale=1.4}]{
\coordinate (t1) at (-4ex,0ex);
\coordinate (t2) at (4ex,0ex);
\draw[thick] (t1) -- (t2);
\filldraw[color=black, fill=black, thick] (t1) circle (0.1ex) node[anchor=south]{\scriptsize{$\tau_1$}};
\filldraw[color=black, fill=black, thick] (t2) circle (0.1ex) node[anchor=south]{\scriptsize{$\tau_2$}};
\node[anchor=north] at (0ex,0ex) {\scriptsize{$k$}};
}
}
\def\externalpropagatorWFU{\tikz[baseline=-0.6ex,scale=1.8, every node/.style={scale=1.4}]{
\coordinate (t1) at (-4ex,0ex);
\coordinate (t2) at (4ex,0ex);
\draw[thick] (t1) -- (t2);
\filldraw[color=black, fill=white, thick] ($(t2) + (-0.4ex,-0.4ex)$) rectangle ($(t2) + (0.4ex,0.4ex)$);
\node[anchor=south] at (t2) {\scriptsize{$\tau$}};
\filldraw[color=black, fill=black, thick] (t1) circle (0.1ex);
\node[anchor=south] at ($(t1) + (0ex,-0.1ex)$){\scriptsize{$\tau_1$}};
\node[anchor=north] at (0ex,0ex) {\scriptsize{$k$}};
}
}
\begin{eqnarray}\label{eq:general-WFU-diagram}
    \nvertexb &\longrightarrow& -(2\pi)^3\delta^{(3)}\qty(\sum_j \bk_j) i^{n + 1}\lambda_n \nn\\
    && \times\int_{-\infty}^\tau \dd{\bar\tau}a^4(\bar\tau)\Big[\dotso\Big]. 
\end{eqnarray}
Inside the square brackets of Eq.~\eqref{eq:general-WFU-diagram}, we need to write every propagator attached to the vertex, which can be either of two types: internal (bulk-to-bulk)
\begin{equation}
\begin{aligned}
\internalpropagatorWFU \longrightarrow G(k; \tau_1, \tau_2),
\end{aligned}
\end{equation}
or external (bulk-to-boundary)
\begin{equation}
\externalpropagatorWFU \longrightarrow -iK(k,\tau_1).
\end{equation}
As in typical Feynman rules, propagators whose momentum is integrated over correspond to loops. Since we are considering single-vertex diagrams, such loops must flow in and out of the same vertex. Finally, symmetry factors work as in the usual Wick theorem.

In analogy with the computation that led to Eq.~\eqref{npt'}, any monomial interaction of order $m\geq n$ in the Taylor expansion in Eq.~\eqref{V-bare-T} will contribute to the $n$-point wavefunction coefficient by dressing the bare coupling $\lambda_n^{\rm b}$ with an arbitrary number of daisy loops.
Then, per the above rules, the wavefunction coefficient with $m$ external legs can be represented via the following series over Witten diagrams:
\def\treediagram{\tikz[baseline=-1.4ex]{
\coordinate (P) at (0,-3ex);
\coordinate (C) at (-9ex,4ex);
\coordinate (k1) at (-8ex,4ex);
\coordinate (k2) at (-4ex,4ex);
\coordinate (k1n) at (4ex,4ex);
\coordinate (kn) at (8ex,4ex);
\draw (C) circle (0.01ex) node[anchor=east]{\footnotesize{$\tau$}};
\draw[thick,dashed] (-9ex,4ex) -- (9ex,4ex);
\draw[thick] (P) -- (k1);
\draw[thick] (P) -- (k2);
\draw[thick] (P) -- (k1n);
\draw[thick] (P) -- (kn);
\filldraw[color=black, fill=black, thick] (-1.2ex,2ex) circle (0.1ex);
\filldraw[color=black, fill=black, thick] (0ex,2ex) circle (0.1ex);
\filldraw[color=black, fill=black, thick] (1.2ex,2ex) circle (0.1ex);
\node at ($(P) + (-0.5ex,0)$) [anchor=east]{\footnotesize{$\lambda^{\text{b}}_{m},\:\bar{\tau}$}};
\filldraw[color=black, fill=white, thick] ($(k1) + (-0.5ex,-0.5ex)$) rectangle ($(k1) + (0.5ex,0.5ex)$) node[anchor=south]{\footnotesize{$\bk_1$}};
\filldraw[color=black, fill=white, thick] ($(k2) + (-0.5ex,-0.5ex)$) rectangle ($(k2) + (0.5ex,0.5ex)$) node[anchor=south]{\footnotesize{$\bk_2$}};
\filldraw[color=black, fill=white, thick] ($(k1n) + (-0.5ex,-0.5ex)$) rectangle ($(k1n) + (0.5ex,0.5ex)$) node[anchor=south]{\footnotesize{$\bk_{m-1}$}};
\filldraw[color=black, fill=white, thick] ($(kn) + (-0.5ex,-0.5ex)$) rectangle ($(kn) + (0.5ex,0.5ex)$) node[anchor=south]{\footnotesize{$\bk_m$}};
\filldraw[color=black, fill=black, thick] (P) circle (0.1ex);
}
}
\def\oneloopdiagram{\tikz[baseline=-1.4ex]{
\coordinate (P) at (0,-3ex);
\coordinate (C) at (-9ex,4ex);
\coordinate (k1) at (-8ex,4ex);
\coordinate (k2) at (-4ex,4ex);
\coordinate (k1n) at (4ex,4ex);
\coordinate (kn) at (8ex,4ex);
\draw (C) circle (0.01ex) node[anchor=east]{\footnotesize{$\tau$}};
\draw[thick,dashed] (-9ex,4ex) -- (9ex,4ex);
\draw[thick] (P) -- (k1);
\draw[thick] (P) -- (k2);
\draw[thick] (P) -- (k1n);
\draw[thick] (P) -- (kn);
\filldraw[color=black, fill=black, thick] (-1.2ex,2ex) circle (0.1ex);
\filldraw[color=black, fill=black, thick] (0ex,2ex) circle (0.1ex);
\filldraw[color=black, fill=black, thick] (1.2ex,2ex) circle (0.1ex);
\node at ($(P) + (-0.5ex,0)$) [anchor=east]{\footnotesize{$\lambda^{\text{b}}_{m+2},\:\bar{\tau}$}};
\filldraw[color=black, fill=white, thick] ($(k1) + (-0.5ex,-0.5ex)$) rectangle ($(k1) + (0.5ex,0.5ex)$) node[anchor=south]{\footnotesize{$\bk_1$}};
\filldraw[color=black, fill=white, thick] ($(k2) + (-0.5ex,-0.5ex)$) rectangle ($(k2) + (0.5ex,0.5ex)$) node[anchor=south]{\footnotesize{$\bk_2$}};
\filldraw[color=black, fill=white, thick] ($(k1n) + (-0.5ex,-0.5ex)$) rectangle ($(k1n) + (0.5ex,0.5ex)$) node[anchor=south]{\footnotesize{$\bk_{m-1}$}};
\filldraw[color=black, fill=white, thick] ($(kn) + (-0.5ex,-0.5ex)$) rectangle ($(kn) + (0.5ex,0.5ex)$) node[anchor=south]{\footnotesize{$\bk_m$}};
\draw[thick,scale=3] (0,-1ex)  to[in=-70,out=-110,loop] (0,-1ex);
\filldraw[color=black, fill=black, thick] (P) circle (0.1ex);
}
}
\def\twoloopdiagram{\tikz[baseline=-1.4ex]{
\coordinate (P) at (0,-3ex);
\coordinate (C) at (-9ex,4ex);
\coordinate (k1) at (-8ex,4ex);
\coordinate (k2) at (-4ex,4ex);
\coordinate (k1n) at (4ex,4ex);
\coordinate (kn) at (8ex,4ex);
\draw (C) circle (0.01ex) node[anchor=east]{\footnotesize{$\tau$}};
\draw[thick,dashed] (-9ex,4ex) -- (9ex,4ex);
\draw[thick] (P) -- (k1);
\draw[thick] (P) -- (k2);
\draw[thick] (P) -- (k1n);
\draw[thick] (P) -- (kn);
\filldraw[color=black, fill=black, thick] (-1.2ex,2ex) circle (0.1ex);
\filldraw[color=black, fill=black, thick] (0ex,2ex) circle (0.1ex);
\filldraw[color=black, fill=black, thick] (1.2ex,2ex) circle (0.1ex);
\node at ($(P) + (-0.5ex,0)$) [anchor=east]{\footnotesize{$\lambda^{\text{b}}_{m+4},\:\bar{\tau}$}};
\filldraw[color=black, fill=white, thick] ($(k1) + (-0.5ex,-0.5ex)$) rectangle ($(k1) + (0.5ex,0.5ex)$) node[anchor=south]{\footnotesize{$\bk_1$}};
\filldraw[color=black, fill=white, thick] ($(k2) + (-0.5ex,-0.5ex)$) rectangle ($(k2) + (0.5ex,0.5ex)$) node[anchor=south]{\footnotesize{$\bk_2$}};
\filldraw[color=black, fill=white, thick] ($(k1n) + (-0.5ex,-0.5ex)$) rectangle ($(k1n) + (0.5ex,0.5ex)$) node[anchor=south]{\footnotesize{$\bk_{m-1}$}};
\filldraw[color=black, fill=white, thick] ($(kn) + (-0.5ex,-0.5ex)$) rectangle ($(kn) + (0.5ex,0.5ex)$) node[anchor=south]{\footnotesize{$\bk_m$}};
\draw[thick,scale=3] (0,-1ex)  to[in=-90,out=-130,loop] (0,-1ex);
\draw[thick,scale=3] (0,-1ex)  to[in=-50,out=-90,loop] (0,-1ex);
\filldraw[color=black, fill=black, thick] (P) circle (0.1ex);
}
}
\begin{widetext}
\begin{flalign}
\psi_m'(\bk_1,\dotsc,\bk_m) &= \treediagram + \oneloopdiagram + \twoloopdiagram + \cdots \notag \\
 &= -i^{m+1}(-1)^m\sum_{L=0}^\infty \frac{\lambda_{m + 2L}^{\text{b}}}{L!2^L}\int_{-\infty}^{\tau}\frac{\dd{\bar{\tau}}}{(H\bar\tau )^4}iK(k_1,\bar{\tau})\dotsb \, iK(k_m,\bar{\tau})\qty[\int_{\bk}G(k;\bar{\tau},\bar{\tau})]^L,
\label{full-diagram}
\end{flalign}
\end{widetext}
with $L$ counting the loops in each diagram.

The loop integrals in the second line of the above equation define the following bulk-to-bulk variance:
\begin{flalign}
\!\!\sigma_{\Psi}^2 &\equiv \!\!\int_{\bk}G(k;\bar{\tau},\bar\tau) \notag \\ 
&= \!\!\frac{H^2}{4\pi^2}\!\!\int\!\!\frac{\dd{p}}{p}\qty[1 \!+\! \qty(\frac{p}{H})^2\!\! -\! \qty(1 + 2i\frac{p}{H} - \qty(\frac{p}{H})^2)e^{-\frac{2ip}{H}}],
\label{eq:sigma_psi-final}
\end{flalign}
where in the second line we changed the integration variable to physical momentum. Even though we will not need the explicit form of $\sigma_{\Psi}^2$ (which is easy to compute), it is worth pointing out that, contrary to the variance in Eq.~\eqref{variance-0}, this expression is regular in the infrared, while it diverges in the ultraviolet. Nevertheless, as we shall see in the next section, the IR divergence kicks back in at the level of the field correlator, which is a crucial feature since the WFU must yield the same results as the other methods. Finally, note that just as in the case of the field variance in Eq.~\eqref{variance-0}, here too, 
passing to physical momentum renders $\sigma_{\Psi}^2$ manifestly time-independent, allowing us to pull it out of the time-integral of Eq.~\eqref{full-diagram}. Consequently, the wavefunction coefficients attain the following form:
\begin{flalign}
\!\!\!\psi_m'(\bk_1,\dotso,\bk_m) =& -i\sum_{L=0}^\infty \frac{\lambda_{m + 2L}^{\text{b}}}{L!}\qty(\frac{\sigma_{\Psi}^2}{2})^L \notag \\
&\times\!\!\! \int_{-\infty}^{\tau}\frac{\dd{\bar{\tau}}}{(H\bar\tau)^4}K(k_1,\bar{\tau})\dotsb K(k_m,\bar{\tau}).
\label{eq:psi_m'-final}
\end{flalign}
The first line corresponds to a sum of bulk loops dressing the bare coupling $\lambda^{\rm b}_m$ much like the sum in Eq.~\eqref{lambda-ren}, with the only difference being the variance. In the next section, we will see how this series combines with a boundary loop resummation to yield Eq.~\eqref{lambda-ren}.

\section{Loop-resumed correlation functions} \label{sec:correlation-functions}

Once the wavefunction $\Psi[\phi]$ is known, one can compute boundary $n$-point correlation functions as
\begin{equation} \label{eq:n-corr-func-definition-0}
\expval{\phi_{\bk_1\dotsb\bk_n}} = \frac{\int\D\phi \, \phi_{\bk_1}\dotsb \,\phi_{\bk_n} \abs{\Psi[\phi]}^2}{\int\D\phi \,\abs{\Psi[\phi]}^2},
\end{equation}
which is nothing but the Born rule. 
Upon inserting the expansion in wavefunction coefficients given in Eq.~\eqref{eq:series-expansion-WFU}, this can be reexpressed to the desired order as
\begin{equation}\label{eq:n-corr-func-definition}
\expval{\phi_{\bk_1\dotsb\bk_n}} = \frac{(2\pi)^{3n}}{Z[0]}\frac{\delta}{i\delta J_{-\bk_1}}\dotsb \frac{\delta}{i\delta J_{-\bk_n}} Z[J]\Bigg|_{J=0},
\end{equation}
where the partition function is defined as
\bea \label{Z-series}
\!\!\!\!Z[J] &\equiv& \exp\Bigg\{\sum_{m=2}^\infty \frac{1}{m!}\int_{\bq_1,\dotso,\bq_m} 2\text{Re}\,\psi_m(\bq_1,\dotso,\bq_m) \nn\\
&\qquad& \times \;(2\pi)^{3m}\frac{\delta}{i\delta J_{-\bq_1}} \dotsb \frac{\delta}{i\delta J_{-\bq_m}}\Bigg\}Z_0[J],
\eea
with the Gaussian part given by
\begin{equation}
Z_0[J] = \exp{-\frac{1}{2}\int_{\bq} J_{\bq} \Delta(q) J_{-\bq}}.
\end{equation}
The free-theory boundary propagator can be easily deduced using the quadratic action in Eq.~\eqref{eq:psi_n-formula}:
\begin{equation} 
\Delta(q)  \equiv \frac{1}{2\text{Re}\,\psi_{2,\rm{free}}'(q)} = \frac{H^2}{2q^3}.
\label{Delta-prop}
\end{equation}

The correlators in Eq.~\eqref{eq:n-corr-func-definition} can be computed via a diagrammatic expansion accompanied by a set of designated Feynman rules which are different than those encountered in the previous section. Here, an $n$-legged vertex represents not an order $n$ monomial in the Taylor expansion of the potential but the $n$-th wavefunction coefficient. The vertex rule thus reads
\def\nvertexbcorrelation{\tikz[baseline=-0.6ex,scale=1.8, every node/.style={scale=1.4}]{
\coordinate (v1) at (0ex,0ex);
\coordinate (phi1) at (-3ex,2ex);
\coordinate (phi2) at (-4ex,0ex);
\coordinate (phi3) at (-3ex,-2ex);
\coordinate (phi4) at (3ex,2ex);
\coordinate (phi5) at (4ex,0ex);
\coordinate (phi6) at (3ex,-2ex);
\draw[thick] (v1) -- (phi1);
\draw[thick] (v1) -- (phi2);
\draw[thick] (v1) -- (phi3);
\draw[thick] (v1) -- (phi4);
\draw[thick] (v1) -- (phi5);
\draw[thick] (v1) -- (phi6);
\filldraw[color=black, fill=black, thick] (v1) circle (0.1ex);
\node[anchor=south] at ($(v1)+(0,-2.5ex)$) {\scriptsize{$\psi_n$}};
\node[anchor=south] at ($(v1)+(0,+1.0ex)$) {\scriptsize{$\cdots$}};
\node[anchor=south] at ($(v1)+(0,+2.0ex)$) {\scriptsize{$n\;\rm{legs}$}};
}
}
\begin{flalign}
\nvertexbcorrelation \rightarrow&  \;(2\pi)^3\delta^{(3)}\qty(\sum_j \bk_j)    \notag \\ & \times\;2\text{Re}\,\psi_n'(\bk_1,\dotso,\bk_n) \Big[\dotso\Big].
\end{flalign}
As before, inside the square brackets we need to include every propagator attached to the vertex.
As already noted, contrary to the diagrams involved in the calculation of the wavefunction coefficient of Eq.~\eqref{full-diagram}, the current diagrams are evaluated entirely at the boundary $\tau$, and hence we do not have to consider time integrals. Moreover, lines now represent boundary propagators, as defined in Eq.~\eqref{Delta-prop}:
\def\internalpropagatorcorrelators{\tikz[baseline=-0.6ex,scale=1.8, every node/.style={scale=1.4}]{
\coordinate (t1) at (-4ex,0ex);
\coordinate (t2) at (4ex,0ex);
\draw[thick] (t1) -- (t2);
\filldraw[color=black, fill=black, thick] (t1) circle (0.1ex);
\filldraw[color=black, fill=black, thick] (t2) circle (0.1ex);
\node[anchor=north] at (0ex,0ex) {\scriptsize{$k$}};
}
}
\begin{equation}
\internalpropagatorcorrelators \longrightarrow \Delta(k).
\end{equation}
Finally, as usual, propagators whose momentum is integrated over correspond to loops, and, since there is only one vertex per diagram (representing a single wavefunction coefficient linear in the couplings $\lambda_n$), these are again daisy loops.

As per these rules, the $n$-point correlation function evaluated at the boundary can be written as the following series:
\def\treelevelcorrelation{\tikz[baseline=-1.4ex]{
\coordinate (P) at (0,0);
\coordinate (C) at (-9ex,4ex);
\coordinate (k1) at (-5ex,-6ex);
\coordinate (k2) at (-5ex,6ex);
\coordinate (k1n) at (5ex,6ex);
\coordinate (kn) at (5ex,-6ex);
\draw[thick] (P) -- (k1);
\draw[thick] (P) -- (k2);
\draw[thick] (P) -- (k1n);
\draw[thick] (P) -- (kn);
\node at (P) [anchor=east]{\footnotesize{$\psi_{n}$}};
\filldraw[color=black, fill=black, thick] (-1.2ex,4ex) circle (0.07ex);
\filldraw[color=black, fill=black, thick] (0ex,4ex) circle (0.07ex);
\filldraw[color=black, fill=black, thick] (1.2ex,4ex) circle (0.07ex);
\filldraw[color=black, fill=black, thick] (-1.2ex,-4ex) circle (0.07ex);
\filldraw[color=black, fill=black, thick] (0ex,-4ex) circle (0.07ex);
\filldraw[color=black, fill=black, thick] (1.2ex,-4ex) circle (0.07ex);
\filldraw[color=black,fill=black] (k1) circle (0.1ex) node[anchor=north]{\footnotesize{$\bk_1$}};
\filldraw[color=black,fill=black] (k2) circle (0.1ex) node[anchor=south]{\footnotesize{$\bk_2$}};
\filldraw[color=black,fill=black] (k1n) circle (0.1ex) node[anchor=south]{\footnotesize{$\bk_{n-1}$}};
\filldraw[color=black,fill=black] (kn) circle (0.1ex) node[anchor=north]{\footnotesize{$\bk_n$}};
}
}
\def\oneloopdiagramcorrelation{\tikz[baseline=-1.4ex]{
\coordinate (P) at (0,0);
\coordinate (C) at (-9ex,4ex);
\coordinate (k1) at (-5ex,-6ex);
\coordinate (k2) at (-5ex,6ex);
\coordinate (k1n) at (5ex,6ex);
\coordinate (kn) at (5ex,-6ex);
\draw[thick] (P) -- (k1);
\draw[thick] (P) -- (k2);
\draw[thick] (P) -- (k1n);
\draw[thick] (P) -- (kn);
\node at (P) [anchor=east]{\footnotesize{$\psi_{n + 2}$}};
\filldraw[color=black, fill=black, thick] (-1.2ex,4ex) circle (0.07ex);
\filldraw[color=black, fill=black, thick] (0ex,4ex) circle (0.07ex);
\filldraw[color=black, fill=black, thick] (1.2ex,4ex) circle (0.07ex);
\filldraw[color=black, fill=black, thick] (-1.2ex,-4ex) circle (0.07ex);
\filldraw[color=black, fill=black, thick] (0ex,-4ex) circle (0.07ex);
\filldraw[color=black, fill=black, thick] (1.2ex,-4ex) circle (0.07ex);
\filldraw[color=black,fill=black] (k1) circle (0.1ex) node[anchor=north]{\footnotesize{$\bk_1$}};
\filldraw[color=black,fill=black] (k2) circle (0.1ex) node[anchor=south]{\footnotesize{$\bk_2$}};
\filldraw[color=black,fill=black] (k1n) circle (0.1ex) node[anchor=south]{\footnotesize{$\bk_{n-1}$}};
\filldraw[color=black,fill=black] (kn) circle (0.1ex) node[anchor=north]{\footnotesize{$\bk_n$}};
\draw[thick,scale=3] (P)  to[in=25,out=-25,loop] (P);
}
}
\def\twoloopdiagramcorrelation{\tikz[baseline=-1.4ex]{
\coordinate (P) at (0,0);
\coordinate (C) at (-9ex,4ex);
\coordinate (k1) at (-5ex,-6ex);
\coordinate (k2) at (-5ex,6ex);
\coordinate (k1n) at (5ex,6ex);
\coordinate (kn) at (5ex,-6ex);
\draw[thick] (P) -- (k1);
\draw[thick] (P) -- (k2);
\draw[thick] (P) -- (k1n);
\draw[thick] (P) -- (kn);
\node at (P) [anchor=east]{\footnotesize{$\psi_{n + 4}$}};
\filldraw[color=black, fill=black, thick] (-1.2ex,4ex) circle (0.07ex);
\filldraw[color=black, fill=black, thick] (0ex,4ex) circle (0.07ex);
\filldraw[color=black, fill=black, thick] (1.2ex,4ex) circle (0.07ex);
\filldraw[color=black, fill=black, thick] (-1.2ex,-4ex) circle (0.07ex);
\filldraw[color=black, fill=black, thick] (0ex,-4ex) circle (0.07ex);
\filldraw[color=black, fill=black, thick] (1.2ex,-4ex) circle (0.07ex);
\filldraw[color=black,fill=black] (k1) circle (0.1ex) node[anchor=north]{\footnotesize{$\bk_1$}};
\filldraw[color=black,fill=black] (k2) circle (0.1ex) node[anchor=south]{\footnotesize{$\bk_2$}};
\filldraw[color=black,fill=black] (k1n) circle (0.1ex) node[anchor=south]{\footnotesize{$\bk_{n-1}$}};
\filldraw[color=black,fill=black] (kn) circle (0.1ex) node[anchor=north]{\footnotesize{$\bk_n$}};
\draw[thick,scale=3] (P)  to[in=50,out=0,loop] (P);
\draw[thick,scale=3] (P)  to[in=0,out=-50,loop] (P);
}
}
\begin{widetext}
\begin{flalign}
    \expval{\phi_{\bk_1\dotsb\bk_n}}' &= \treelevelcorrelation + \;\oneloopdiagramcorrelation +\; \twoloopdiagramcorrelation + \dotso \notag \\
    &= \sum_{L=0}^\infty \frac{1}{L!\;2^L}\int_{\bq_1}\dotsb\int_{\bq_L} 2\; \text{Re}\,\psi_{n+2L}'(\bk_1,\dotsc,\bk_n,\bq_1,-\bq_1,\dotsc,\bq_{L},-\bq_L) \Delta(k_1)\dotsb \Delta(k_n)\Delta(q_1)\dotsb \Delta(q_L).
\label{n-point-in-function-of-WFU}
\end{flalign}
\end{widetext}
Comparing this expression to Eq.~\eqref{full-diagram}, we immediately see that it is not a power series in loops since now the wavefunction coefficients depend on the internal momenta running in the loops, which are being integrated over.  
To reach a genuine power series, the final step is to insert the wavefunction coefficient $\psi'_{n+2L}$ from Eq.~\eqref{eq:psi_m'-final} into Eq.~\eqref{n-point-in-function-of-WFU}, which yields a factorizable form:
\begin{widetext}
\begin{flalign}
\expval{\phi_{\bk_1\dotsb\bk_n}}' &= -\sum_{L_2=0}^\infty\frac{1}{L_2!\;2^{L_2}}\int_{\bq_1}\dotsb\int_{\bq_{L_2}}\Delta(k_1)\dotsb\Delta(k_n)\Delta(q_1)\dotsb\Delta(q_{L_2}) \notag \\
&\quad \times 2\;\text{Re}\Bigg\{ i\sum_{L_1=0}^\infty \frac{\lambda_{n + 2(L_1 + L_2)}^{\text{b}}}{L_1!}\qty(\frac{\sigma_{\Psi}^2}{2})^{L_1}\int_{-\infty}^\tau \frac{\dd{\bar{\tau}}}{(H\bar\tau)^4}K(k_1,\bar\tau)\dotsb K(k_n,\bar{\tau})K^2(q_1,\bar{\tau})\dotsb K^2(q_{L_2},\bar{\tau}) \Bigg\} \notag\\
&= 2\;\text{Im}\Bigg\{ \int_{-\infty}^\tau\frac{\dd{\bar{\tau}}}{(H\bar\tau)^4}K(k_1,\bar{\tau})\Delta(k_1)\dotsb K(k_n,\bar{\tau})\Delta(k_n) \notag \\
&\qquad\qquad \times \sum_{L_1=0}^\infty  \frac{\lambda_{n + 2(L_1 + L_2)}^{\text{b}}}{L_1!}\qty(\frac{\sigma_{\Psi}^2}{2})^{L_1}   \sum_{L_2=0}^\infty \frac{1}{L_2!\;2^{L_2}} \qty[\int_{\bq}K^2(q,\bar{\tau})\Delta(q)]^{L_2}\Bigg\}.
\label{eq:n-function-1}
\end{flalign}
\end{widetext}

The loop integrals of the above equation define a bulk-to-boundary variance:
\begin{flalign}
\sigma_{\Phi}^2 &\equiv \int_{\bq}K^2(q,\bar\tau)\Delta(q) \notag \\
&= \frac{H^2}{4\pi^2}\int\frac{\dd{p}}{p}\qty(1 + 2i\frac{p}{H} - \qty(\frac{p}{H})^2)e^{-\frac{2ip}{H}},
\label{eq:sigma_phi-final}
\end{flalign}
which is again manifestly time-independent upon switching to physical momentum as in the second line. Unlike the bulk-to-bulk variance in Eq.~\eqref{eq:sigma_psi-final}, this variance diverges logarithmically in the IR. Further, noting that 
\begin{equation}
\begin{aligned}
K(k,\bar{\tau})\Delta(k) = g(k,\tau,\bar{\tau}), 
\end{aligned}
\end{equation}
with $g(k,\tau,\bar{\tau})$ given by Eq.~\eqref{G-uu},
and grouping the sums into 
\begin{equation} \label{lambda-obs}
\!\!\!\!\bar\lambda_n= \sum_{L_2=0}^\infty \sum_{L_1=0}^\infty \frac{1}{L_1!\;L_2!}\qty(\frac{\sigma_{\Psi}^2}{2})^{L_1}\qty(\frac{\sigma_{\Phi}^2}{2})^{L_2}\lambda_{n + 2(L_1+L_2)}^{\text{b}},
\end{equation}
we can finally rewrite Eq.~\eqref{eq:n-function-1} as
\begin{equation}\label{eq:<...>-final-WFU}
\expval{\phi_{\bk_1\dotsb\bk_n}}' = \frac{\bar\lambda_n}{H^4}2\,\text{Im}\Bigg\{  \int_{-\infty}^{\tau}\frac{\dd{\bar\tau}}{{\bar\tau}^4}g(k_1,\tau,\bar{\tau})\dotsb g(k_n,\tau,\bar{\tau})\Bigg\}.
\end{equation}
Next, we may use the binomial theorem to simplify Eq.~\eqref{lambda-obs} to
\begin{equation} \label{lambda-obs-ren}
\bar\lambda_n = \sum_{L=0}^\infty \frac{1}{L!}\qty(\frac{\sigma_{\Psi}^2 + \sigma_{\Phi}^2}{2})^L \lambda_{n + 2L}^{\text{b}}.
\end{equation}
Note that Eqs.~\eqref{eq:sigma_psi-final} and \eqref{eq:sigma_phi-final} imply
\begin{equation}
    \sigma_{\Psi}^2 + \sigma_{\Phi}^2 = \sigma_{\rm tot}^2,
\end{equation}
with $\sigma_{\rm tot}^2$ as in Eq.~\eqref{variance-0-1}. This renders $\bar\lambda_n=\lambda_n^{\rm obs}$ of Eq.~\eqref{lambda-ren}. In other words, we have reached the same loop-resumed correlation function as the one written in Eq.~\eqref{npt'} calculated with the SK path integral. 

\section{Loop-resumed wavefunction coefficients}
Having established the equivalence of the two methods for arbitrary single-vertex diagrams, let us go back to the wavefunction coefficients and see how the resummation works at this level. Upon inverting the relation in Eq.~\eqref{lambda-obs-ren} using the differential structure of the Weierstrass transform, we obtain $\lambda_n^{\text{b}} = \sum_{L=0}^\infty \frac{1}{L!}(-1)^L\qty(\frac{\sigma_{\rm tot}^2}{2})^L \bar\lambda_{n + 2L}.$ 
Substituting this into Eq.~\eqref{eq:psi_m'-final}, we find
\begin{equation}
\!\!\!\psi_m'(\bk_1,\dotso,\bk_m) =-i\Lambda_{m}
 \!\!\! \int_{-\infty}^{\tau}\!\!\frac{\dd{\bar{\tau}}}{(H\bar\tau)^4}K(k_1,\bar{\tau})\dotsb K(k_m,\bar{\tau}),
\label{eq:psi_m'-final-2}
\end{equation}
with 
\begin{flalign} \label{Lambda}
\Lambda_{m} \equiv \sum_{L = 0}^{\infty}\frac{(-1)^{L}}{L!}\qty(\frac{\sigma_{\Phi}^2}{2})^{L} \lambda^{\rm obs}_{m + 2L},
\end{flalign}
and $\sigma_{\Phi}^2$ given by Eq.~\eqref{eq:sigma_phi-final}. 

Note that now the IR behavior of the wavefunction coefficients is in accordance with that of the correlators: they both diverge logarithmically in the IR. The wavefunction coupling constants $\Lambda_{m}$ cannot be renormalized via the introduction of counterterms~\cite{Huenupi:2024ksc} in the diagrammatic expansion of Eq.~\eqref{full-diagram} since, here, $\lambda^{\rm obs}$ is the observable, finite coupling constant. However, this is not a problem since $\Lambda_{m}$ is not observable. 
As a last step, we may 
use the differential representation of the Weierstrass transform to recognize $\Lambda_m$ as the Taylor coefficients of a complex potential. (Note that the variance $\sigma_\Phi^2$ is complex). Following the procedure that led us to Eqs.~\eqref{Vobs-def} and \eqref{Vobs-W}, we define the following potential:
\begin{flalign}
e^{\frac{\sigma_\Psi^2}{2} \frac{\partial^2}{\partial \Phi^2} } {\cal V}_{\rm b}(\Phi) \equiv {\cal V}_\psi(\Phi) = \sum_{m = 0}^{\infty}\Lambda_{m} \frac{\Phi^m}{m!}.
\label{eq:psi_m'-final-3}
\end{flalign}
Equation~\eqref{eq:psi_m'-final-2} thus informs us that single-vertex, loop resumed wavefunction coefficients emanating from a potential ${\cal V}_{\rm b}(\Phi)$ are equivalent to tree-level  wavefunction coefficients computed using ${\cal V}_\psi(\Phi)$. Moreover, due to the relation in Eq.~\eqref{Lambda}, these wavefunction coefficients yield the loop-resumed correlators in Eq.~\eqref{npt'} whose amplitudes are set by the finite couplings $\lambda^{\rm obs}_n$.

\section{Conclusions}

It is well-known that, at tree or 1-loop levels, $n$-point correlation functions for an interacting scalar field in de Sitter space can be calculated using either 
the Schwinger-Keldysh path integral (equivalently the in-in formalism) or the WFU method~\cite{Lee:2023jby,Creminelli2024}. 
We have demonstrated that this equivalence persists, as expected, in an exact manner for $n$-point correlators sourced by an arbitrary analytic potential. 

At the perturbative level, the arbitrariness of the interaction translates into an infinite number of vertices, which inevitably decorate each contact diagram with an infinite number of loops. The crucial observation that allows for a complete match to linear order in the potential is that the bulk loops of the wavefunction coefficients and the boundary loops of the correlators combine to yield an all-loop resumed, renormalized potential with coupling constants given by Eq.~\eqref{lambda-ren}. In addition, we have shown how to resum bulk loops at the level of the wavefunction coefficients. At tree level, a single wavefunction coefficient corresponds to a single correlator; at loop level the map is not one-to-one. Equation~\eqref{eq:psi_m'-final-2} shows exactly how to compute wavefunction coefficients using the tree-level vertices $\Lambda_m$, hence restoring the one-to-one correspondence between loop-resumed wavefunction coefficients and loop-resumed correlators.

Our result also advocates for the use of physical variables, which in some cases render integrals manifestly invariant under time translations. This is a property that should persist in diagrams beyond the single-vertex family discussed here and whose study we leave for future work~\cite{PST}.

The loop-resumed wavefunction may be directly applicable to the bootstrap program, where it can be used to extend the computation of correlators beyond tree level. This, in turn, can serve as a bridge between cosmological dynamics and (for instance) axion phenomenology, since the latter involves nonperturbative trigonometric potentials~\cite{Marsh:2015xka} which fall in the category considered herein. 

\begin{acknowledgments}

We wish to thank Gabriel Mar\'in Mac\^edo, Sebasti\'an C\'espedes, and Nicol\'as Parra for comments. GAP and SS acknowledge support from the Fondecyt Regular projects 1210876 and 1251511 (ANID). JH is supported by ANID-Subdirección del Capital Humano/Magíster Nacional/2023-22231422. EH is supported by the MIT Department of Physics.

\end{acknowledgments}

\bibliography{bibliography.bib}

\end{document}